\documentstyle[11pt,citesort]{article}
\def \mm {{\rm m}}
\def \la {\label}
\def\ov{{\mbox{\raisebox{2pt}{\large $ \!\! /\!\!$}}}}

\def\del{\partial}
\def \a {\alpha}
\def \b {\beta}

\def\th{\theta}
\def \ha {{1 \over 2}}

\def \thetat {\tilde{\theta}}

\def \I {i}
\def \is {\! & \! = \! & \! }
\newcommand{\newsubsection}[1]{
\vspace{1cm}
\pagebreak[3]
\addtocounter{subsection}{1}
\addcontentsline{toc}{subsection}{\protect
\numberline{
\arabic{subsection}}{#1}}
\noindent{\large\bf 
#1}
\nopagebreak
\vspace{2mm}
\nopagebreak}

\newlength{\extraspace}
\newlength{\extraspaces}
\newcommand{\ba}{\begin{eqnarray}
\addtolength{\abovedisplayskip}{\extraspaces}
\addtolength{\belowdisplayskip}{\extraspaces}
\addtolength{\abovedisplayshortskip}{\extraspace}
\addtolength{\belowdisplayshortskip}{\extraspace}}
\newcommand{\ea}{\end{eqnarray}}
\newcommand{\be}{\begin{equation}
\addtolength{\abovedisplayskip}{\extraspaces}
\addtolength{\belowdisplayskip}{\extraspaces}
\addtolength{\abovedisplayshortskip}{\extraspace}
\addtolength{\belowdisplayshortskip}{\extraspace}}
\newcommand{\ee}{\end{equation}}

\begin{document}
\addtolength{\baselineskip}{.8mm}
\def\calN{{\cal N}}
\def\Box{\square}
\def \Tr{\mbox{Tr\,}}
\def \tr{\mbox{Tr\,}}
\def \a{\alpha}
\def \da{{\dot a}}
\def \db{{\dot b}}
\def \dc{{\dot c}}
\def \dd{{\dot d}}
\def \CH{\mathfrak {CH}}
\def \b{\beta}
\def \vt{\vartheta}
\def \ttheta{{\tilde \theta}}
\def \tSigma{{\tilde \Sigma}}
\def\ignorethis#1{}
\def\ar#1#2{\begin{array}{#1}#2\end{array}}
\def\bear{\begin{eqnarray}}
\def\eear{\end{eqnarray}}
\def\p{\partial }
\def\IR{{\bf R}}

\def \two {{{}_{2}}}

\newcommand{\ggt}{\!>\!\!>\!}
\newcommand{\llt}{\!<\!\!<\!}


\begin{titlepage}
\begin{center}

{\hbox to\hsize{
\hfill PUPT-2040}}
{\hbox to\hsize{
\hfill hep-th/0206059}}

\bigskip

\vspace{6\baselineskip}

{\LARGE \bf B}{\Large \bf its,}
{\LARGE \bf M}{\Large \bf atrices and} {\LARGE 1 $\ov$ \bf N}
\bigskip
\bigskip
\bigskip

{\large Herman Verlinde}\\[1cm]

{ \it Physics Department, Princeton University, Princeton, NJ 08544}\\[5mm]

\vspace*{1.5cm}

{\bf Abstract}\\
\end{center}
\noindent
We propose a simple string bit formalism for interacting strings
in a plane wave background, in terms of supersymmetric quantum
mechanics with a symmetric product target-space. We construct the light-cone
supersymmetry generators and Hamiltonian at finite string coupling. We find
a precise match between string amplitudes and the non-planar
corrections to the correlation functions of BMN operators computed from gauge
theory, and conjecture that this correspondence extends to
all orders in perturbation theory. We also give a simple RG explanation for
why the effective string coupling is $g_\two = {J^2/ N}$ instead
of $g_s = g_{ym}^2$.

\end{titlepage}

\newpage

\newsubsection{Introduction}

Recently a new framework to describe type IIB string
dynamics from ${\cal N}=4$ supersymmetric Yang-Mills theory has been proposed
in \cite{bmn}. It was argued that for states with large R-charge $J$, the
gauge theory in effect amounts to a (discrete) light cone quantization
of strings moving in the background of a pp-wave geometry \cite{bla}
\be
\label{ppwave}
{ds^2 = -4 dx^+ dx^- - \mu^2 (\vec r^2 +\vec y^2) (dx^+)^2
+ d\vec y^2 + d \vec r^2,
\ \quad \ F_{+1234}=F_{+5678}=\frac{\mu}{4\pi^3 g_s\alpha^{'2}}.}
\ee
The specific dictionary proposed in \cite{bmn} involves a one-to-one
identification of a class of single trace operators in the gauge theory
with single string states in this background. At the non-interacting
string level, at leading order in the large $N$ expansion, supporting
evidence was found via the computation of the anomalous scale dimensions of
the gauge theory operators, confirming the identification
\be
\label{lcm}
P_- = \Delta -J.
\ee
Subsequent work has extended this correspondence to low orders in
string perturbation theory.

In this note, we will propose a simple effective description of the
string dynamics in the pp-wave background, in terms of a string bit
language \cite{thornbits} inspired by the formalism of \cite{bmn} as well as by
matrix string theory \cite{lubos}\cite{dvv}.
This reduced description is based on a
simple supersymmetric quantum mechanical model with a symmetric
product target space. We will present evidence that this effective
description exactly reproduces the {\it complete} perturbation
expansion of the ${\cal N}=4$ SYM theory in the BMN limit. In particular,
we will find a precise match between the amplitudes of the string bit
model and the three-point proposed in \cite{seven} and one-loop amplitudes
computed in \cite{four} and \cite{seven}. The philosophy of our approach, h
owever, is to compare the gauge and string theory at a more microscopic level,
by matching both systems directly via their light-cone Hamiltonian
evolution.

The light-cone gauge worldsheet Lagrangian of a string propagating in the
background (\ref{ppwave}) takes the quadratic form
\be\la{lang}
 {\cal L}=
\ha(  \del_+ x_{\I} \del_- x_{\I}  -  \mm^2  x^2_{\I})  +
{\rm i}  ( \theta_a \partial_+  \theta_a  +
\thetat_a \partial_-  \thetat_a -2 \mm    \thetat_a
 \Pi_{ab} \theta_b ) \, ,
\ee
with $\mm = \mu \, p_+$ and  $\Pi = \gamma_{1234}$. The quantization and
symmetries of this Lagrangian have been studied in detail in \cite{mets}.
In accordance with the symmetries of the pp-wave background, it exhibits
a maximally extended sypersymmetry (for all details, see \cite{mets}).

In the following, our aim is to construct an {\it interacting} version
of this string theory, with a discretized world-sheet and with the same
space-time symmetries. The only generators that are expected to receive
corrections at finite string coupling are the light-cone generators
$Q^-$ and $P^-$. Hence, the part of the supersymmetry algebra that will
be most important to us, is
\footnote{Here, relative to the notation in \cite{mets},
$Q = Q^{-1}$ and $\widetilde{Q}=Q^{-2}$.}
\be
\label{lcsym}
\{ Q^\da, {Q}^\db\}
+ \{\widetilde{Q}^\da,\widetilde{Q}^\db\}
=\; \delta^{\da\db} \, H \; +\; \mm
\, J^{\da\db}\, , \qquad \quad [H, Q_\da] = 0
\ee
where $H\equiv P^-$ and $J^{\da\db}$ is a suitable contraction of
gamma matrices with the $SO(4)\!\times\! SO(4)$ Lorentz generators $J^{ij}$,
see \cite{mets}.


A convenient formalism for describing the full multi-string Hilbert space,
is to use a single orbifold field theory on the symmetric product
target space
\be
{\rm Sym}^J {\cal M} = {{\cal M}^J / S_J}
\ee
with $S_J$ the permutation group and $J$ the total DLCQ momentum. In the
case of IIA string theory in flat space, this formalism naturally arises
as the IR limit of matrix string theory
\cite{lubos} \cite{dvv}.\footnote{For some recent work on matrix (string) theory
in pp-wave backgrounds, see \cite{Dasgupta}\cite{related}\cite{rajesh}.}
Moreover, it allows for a remarkably compact description
of the string splitting and joining, via the interaction vertex \cite{dvv}
\be
\label{intee}
g_s V_{int} =  {g_s \over J^2}\sum_{n<m} \int \! d\sigma \,
\Bigl(\tau^i \tilde\tau^j \otimes \Sigma^i \widetilde\Sigma^j\Bigl)_{nm}\, ,
\ee
where each term on the right-hand side represents a suitable twist-field
that implements a simple permutation $(nm)$, interchanging the
$n$-th and $m$-th copy of the target space. This description of matrix string
interactions will be useful to keep in mind in what follows.

\newsubsection{Bits}

Motivated by the BMN formalism, we will now set up the string bit language,
by introducing $J$ copies of supersymmetric phase space coordinates
$\{ p_n^i, x_n^i, \theta_n^a, \thetat_n^a\}$, with $n=1, \ldots, J$,
satisfying canonical commutation relations
\be
[\, p^i_n , x^j_m\, ]  =  \delta^{ij} \delta_{mn} \, , \qquad \ \
\{ \theta^a_n, \theta^b_m \}  =  \delta^{ab} \delta_{mn} \, , \qquad \ \
\{ \thetat^a_n, \thetat^b_m \}  =  \delta^{ab} \delta_{mn}.
\ee
Following the example of matrix string theory, we consider these $J$
copies as obtained via the quantization
of the $J$-th symmetric product ${\rm Sym}_J {\cal M}$ 
of the plane wave target space ${\cal M}$. In other words, we divide
out the symmetric group $S_J$, acting via permutation of on the labels
$n$, thus defining a quantum mechanical orbifold model.

In the context of 2-d field theory, it is well known that the
Hilbert space of such an orbifold field theory decomposes as a direct
sum over twisted sectors. Although the occurrence of such twisted sectors
is often thought of as a purely ``stringy'' phenomenon, they do in fact
also arise in point-particle quantum mechanics. For a clear discussion of this
in the context of cohomology of symmetric orbifold spaces, see \cite{vw}, pp
56-57.
Following the prescription outlined in \cite{vw}, we construct our
Hilbert space as a direct sum of ``twisted sectors''
\be
{\cal H} = \bigoplus_\gamma \, {\cal H}_\gamma
\ee
labeled by conjugacy classes $\gamma$ of the symmetric group $S_J$.
Each twisted sector ${\cal H}_\gamma$ can be thought of as made up from
states localized at the fixed point set of $\gamma$.\ \footnote{Here and in
the following, we use the notation $\gamma$ both for the conjucacy class
as for a specific representative in this class.}
This fixed point set is mapped onto itself by the stabilizer subgroup $N_\gamma$
of permutations $\sigma$ that commute with $\gamma$. Correspondingly, in
${\cal H}_\gamma$, we can act with arbitrary operators
${\cal O}(p,x,\theta)$ that are left {\it invariant} under the action
of $N_\gamma$:
\be
\qquad \qquad \{ \,p^i_n,  x^\I_n, \theta^{a}_n\}\; \rightarrow
\ \{\, p^i_{\sigma(n)}, x^\I_{\sigma(n)},
\theta^{a}_{\sigma(n)}\}\, , \qquad \ \ \sigma \in N_\gamma.
\ee
By copying (with only slight modification) the discussion in \cite{dvv} (on
pp 4-5), it is readily verified that the resulting Hilbert space takes
the form of a sum over multi-string Hilbert spaces, each string with
a discretized worldsheet consisting of $J_\ell$ bits with
$\sum_\ell J_\ell =J$.
As in matrix string theory, we interpret $J_\ell$ as the discrete light-cone
momentum of the string.
The above invariance under the stabilizer subgroup imposes the
constraint
\be
\label{constraint}
(\, U_\ell)^{J_\ell} = 1 \ \, \qquad \quad
U_\ell = e^{2\pi i (L^{(\ell)}_0 - \widetilde{L}^{(\ell)}_0)/J_\ell}
\ee
on each string, with $U_\ell$ the operator that translates the
string bit $x_n$ by one unit. Here the adjacency of two sites is
specified by the twisted boundary condition $\gamma$. So in particular
the ``overall'' translation operator $U = \otimes_{\strut \! \! \ell} \,
U_\ell$ is defined to act via
\be
U X_n U = X_{\gamma(n)}
\ee
with $X_n \! = \! \{p^i_n,x^i_n,\theta^a_n \}$.
It translates the bits only within each individual string, i.e. it does not
translate any bit from one string to the next.
Finally, we notice that each multi-string state is required to be
(anti-)symmetric (according to statistics) under interchange
of strings with equal discrete light-cone momentum $J_\ell$.


The light-cone supersymmetry generators and Hamiltonian of
the free string theory read
\be
\label{expa}
Q_0 = Q_\da^{{}^{\! (0)}} \! + \lambda   \, Q_\da^{{}^{\! (1)}} \ , \qquad
\ \
H_0 = H^{{}^{\! (0)}} \! + \lambda \, H^{{}^{\! (1)}} \!
+ \lambda^2  H^{{}^{\! (2)}}
\ee
with
\ba
\label{lcgen}
Q^{{}^{\! (0)}}_\da \!\!
\is \! \sum_n (p_{n}^{\, i} {\gamma}_i
\, \theta_n \! -  x^{i}_n({\gamma}_i
\Pi)
\thetat_n)
\, , \qquad \quad Q_\da^{{}^{\! (1)}} \!
= \!  \sum_n (x^i_{\gamma(n)}\!\! - x^i_{n})
{\gamma}_i
\, \theta_n  \qquad 
\\[3.5mm]
H^{{}^{\! (0)}} \! \! \is \! \sum_n (  \frac{1}{2 } (p_{i,n}^2 +
 x_{i,n}^2) + 2{\rm i}\, \thetat_n\Pi
\theta_n) \, , \\[3.5mm]
H^{{}^{\! (1)}}\!\! \is\! - \sum_n
{\rm i}(\theta_n  \th_{\gamma(n)}
-\thetat_n\thetat_{\gamma(n)} ) \, , \qquad \quad \ \ \
H^{{}^{\! (2)}}\!
=  \sum_n \frac{1}{2 } (x^{\I}_{\gamma(n)} \!\! - x^i_n)^2
\ea
These expressions are a straightforward discretization of the expressions
in \cite{mets}. Here $\lambda$ is a parameter that controls the size of
the string bits relative to the mass scale $\mm$ in the light-cone
worldsheet Lagrangian (\ref{lang}).

Via \cite{bmn} and \cite{grossea}, there now exists convincing evidence
that the above free light-cone Hamiltonian exactly summarizes the
propagation of the BMN operators in the leading order large N limit,
via the identification (\ref{lcm}) and of $\lambda$ with the 't Hooft
coupling
\be
\lambda^2 = {g_{ym}^2 N}.
\ee
Our next goal is to extend this correspondence to include
string splitting and joining.

\newsubsection{Interactions}

We now add interaction terms to the light-cone generators, as follows
\be
Q^\da = Q_0^\da + g_s S_1^\da
\, , \qquad \quad \ \ H \, =  H_0 + g_\two V_1 + g_\two^2 \; V_2,
\ee
with $g_\two$ the (effective) string coupling. Imposing the light-cone
supersymmetry algebra (\ref{lcsym}) produces the relations
\ba
\label{een}
\{Q_0^{\da},  {S}_1^{\db}\} + \{\widetilde{Q}_0^{\da}, \widetilde{S}_1^{\db} \}   +
\mbox{\footnotesize $(\da \leftrightarrow \db)$}
\is
\delta^{\da\db} \, V_1 \\[3mm]
\label{twee}
[\, H_0 , S_1^{\da}\, ] \, + \, [\, V_1, Q_0^{\da}\,] \is 0  \\[3mm]
\label{drie}
\{ S_1^\da, {S}_1^\db\} \, + \,
\{\widetilde{S}_1^\da,\widetilde{S}_1^\db\} \is \delta^{\da\db} \, V_2 \\[3mm]
\label{vier}
[\, V_2, Q_0^\da\, ] \, +\,  [\, V_1,S_1^\da\, ] \is 0 .
\ea
We wish to solve these relations via suitable interaction terms $S_1$ and $V_1$
that induce the splitting and joining of strings. It will be a non-trivial
result that the above algebra can  be satisfied, without the need of introducing
any higher order interaction terms than $V_2$.

Following the matrix string theory example, it is natural to look for
operators analogous to the twist-field interaction (\ref{intee}).
Consider operators $\Sigma_{mn}$ that implement a simple transposition
of two string bits via
\be
\label{sigma1}
\qquad \qquad \Sigma_{nm} X_m = X_n
\, \Sigma_{nm}\, , \qquad \qquad
\Sigma_{nm}   X_k = X_k \, \Sigma_{nm} \, \qquad
\mbox{\footnotesize $k \neq m,n$}
\ee
with $X_n \! = \! \{p^i_n,x^i_n,\theta^a_n \}$. Clearly, by acting
with $\Sigma_{mn}$ on a given multi-string sector, we get a different multi-string
sector via
\be
\label{sigma2}
\Sigma_{mn}: \quad {\cal H}_\gamma \rightarrow {\cal H}_{\tilde\gamma}\, ,
\qquad \quad  \mbox{\footnotesize{with \ $\tilde{\gamma} = \gamma \!\circ\! (mn)$}}.
\ee
Depending on whether the two sites $n$ and $m$ in the sector $\gamma$
correspond to one single string (say of length $J_0$) or two separate ones
(say of length $J_1$ and $J_2$),
the new sector  $\tilde{\gamma}$ is obtained by either splitting the single
string in two pieces of length $(m\!- n)$ and $(J_0\! -m\! +n)$, or by
joining the two strings to one of length $J_1\!+J_2$.

Let us now introduce the $S_J$ invariant operator
\be
\label{sigma}
\Sigma = {1\over J^2}\sum_{n<m} \Sigma_{nm}
\ee
and define the interaction terms via
\be
\label{def}
S_1^\da = [ \, Q^\da_0, \Sigma\, ] \, , \ \ \qquad \
V_1 = [\, H_0, \Sigma \, ].
\ee
This form of the interaction term has several motivations. First,
it manifestly solves the equations (\ref{een}) and (\ref{twee}) imposed
by the space-time supersymmetry algebra. Secondly, by the preceding
discussion, it represents a simple splitting and joining interaction.
Finally, the matrix string theory vertex
(\ref{intee}), light-cone string field theory \cite{nastamarc}
and the ${\cal N}=4$ gauge theory calculations \cite{seven}, all
suggest that the interaction vertex should contain, besides the
simple splitting and joining operator, a prefactor quadratic in the string
oscillators. Moreover, we expect that the interaction term should vanish
when $\lambda=0$, i.e. in the free SYM theory. (We will further motivate
this requirement in the next subsection.) Notice that,
since the leading order generators $Q^{{}^{\!(0)}}$ and $H^{{}^{\!(0)}}$ are
permutation invariant, both interaction terms are indeed (at least)
of order $\lambda$.

It is straightforward, via (\ref{sigma1}) and the definition of the light-cone
generators, to explicitly compute $S_1$ and $V_1$. For $S_1$ we find
\be
S_1 = {1\over J^2}
\sum_{m,n} \Sigma_{mn}\Bigl(\theta_m \gamma^i \! (x^i_{\gamma(m)}\!\!-
\!x^i_{\gamma(n)})
+ \theta_{\gamma^{{}^{\! -1}}(m)}\gamma^i\! (x^i_n\! -\! x^i_m)
 -  \delta_{n\gamma(m)}
(\theta_m \gamma^i x^i_n \!-\theta_n \gamma^i x^i_m)\Bigr)
\ee
From this explicit form, it is directly verified
that the $S_1$ terms in fact satisfy a commutation relation of the form
(\ref{drie}), which can thus be taken as the definition of the second order
interaction term $V_2$. The relation (\ref{vier}) is then automatically
satisfied.

Thus the interacting string-bit theory has all the required symmetries,
as well as all required dynamic properties (at least all obvious ones).
We will now make a comparison with the gauge theory amplitudes.

\newsubsection{Matrices}

The interacting string bit theory is constructed to correspond
with the non-planar gauge theory amplitudes in the BMN limit, via the
identification of the effective string coupling $g_\two$ with
\be
g_\two = {J^2 / N},
\ee
and of the light-cone energy with scale dimension $\Delta-J$.
In view of this last correspondence, it seems appropriate to view
the gauge theory amplitudes as obtained via radial evolution
starting from an initial state defined at some given point, say at $r=0$,
which then evolves via the dilation operator to a final state defined
at $r=\infty$. Preferably, from a given initial operator at $r=0$, one
would then like to be able to keep track of how the index structure of the
state evolves with $r$, so that, via an appropriate dictionary,
we can identify the number of strings propagating at this instant and
make a precise comparison with the corresponding light-cone string
diagram.

A technical fact, that seems to be at odds with this philosophy,
is that the two-point functions of single trace operators
in the gauge theory in fact have a non-trivial $1/N$ expansion even at
$g_{ym}\!\! =0$, \cite{four}\cite{bn} and \cite{seven}. This would
seem to indicate that
strings can split and join even without any non-trivial SYM interaction
taking place, thus preventing a precise determination of the number
of strings at given light-cone time $r$. Our point of view, however, is that
the $1/N$ corrections that are present in the free SYM theory, do {\it not}
correspond to proper string interactions, but need to be absorbed into
a redefinition of the precise operator dictionary (using techniques
similar to \cite{cjr}).\footnote{This redefinition is possible, because
string S-matrix elements between in- and out-states, via the above
prescription, in fact correspond to {\it two-point} functions in SYM theory
of corresponding multi-trace operators placed at $r=0$ and $r=\infty$,
rather than multi-point functions. In particular,
the three-point function is an S-matrix element between an one-string and a
two-string state (see below), and {\it not} equal to a three-point function
of the SYM theory.}
This point of view was also taken in \cite{seven}.

Modulo this subtlety, single trace operators in first order correspond
to single strings. The cyclicity of the trace then naturally implements
the $L_0-\widetilde{L}_0$ constraint (\ref{constraint}), while via permutations
of the operators in a single or multi-trace expression, we can (similarly
as discussed above) induce string splitting or joining. Combined with the
original analysis of \cite{bmn}, it seems natural to look for a combinatoric
language by which we can characterize the SYM interactions as a composition
of a small number of elementary operations, such as nearest neighbor hopping
or simple permutations.
The general dictionary relies on a one-to-one correspondence between
interaction terms in the SYM and string bit Hamiltonian, via
\ba
\Tr(\theta [Z,\theta]) & \leftrightarrow & H^{{}^{\! (1)}}\!\! + g_\two
[H^{{}^{\! (1)}}, \Sigma] \\[3mm]
\Tr |[Z,X]|^2 & \leftrightarrow  & H^{{}^{\! (2)}} \!\! + g_\two
[ H^{{}^{\! (2)}},\Sigma]  + g_\two^2 V_2^{{}^{\! (2)}}\, ,
\ea
with $V_2 = \lambda^2 V_2^{{}^{\! (2)}}$.
We leave a more detailed check of this dictionary to a future publication.

Instead, to illustrate the correspondence, let us compare the predictions
for some low order amplitudes with existing gauge theory calculations.
This task turns out to be almost trivial, since many of the computations
done in \cite{four}
\cite{seven} can be readily transferred to the string bit language.

First, consider the elementary three point
interaction between an initial single string state $|\, i\, \rangle$,
splitting into a two-string final state $|f_1,f_2\rangle$.
Quite generally, since $V_1 = [H, \Sigma]$, we have that
\be
\label{sevensplit}
\langle \, i\, | V_1 | \,f_1,f_2  \rangle =
(\Delta_i \! - \Delta_{f_1}\! -\Delta_{f_2})\,
\langle\, i\, | \Sigma |\, f_1,f_2  \rangle
\ee
The reduced matrix element on the right-hand side corresponds
to the three-point function computed in the {\it free} SYM gauge theory.
Equation (\ref{sevensplit}) is therefore in direct accordance with the
three-point string vertex proposed in \cite{seven}.

Let us focus on the same class of operators considered in \cite{four}
\cite{seven}
\be
O^J_p = \mbox{\small{${1\over \sqrt{JN^{J+2}}} \sum\limits_{l=1}^J e^{2\pi i l/J}{\rm Tr}
(\phi Z^l \psi Z^{J-l})$}} \
\; \  \leftrightarrow \; \ \ \ O^J_p =
{1\over J^{3/ 2}}\sum\limits_{k,l=1}^J a_k^\dagger \, b_l^\dagger
\, e^{2\pi i p\, (l-k)/J}
\ee
and compute the three point function
\be
{}_{\strut J}\! \langle \, O_p^J\, |\,   V_1 | \,
O_q^{J_1}\,\rangle_{\strut{\! J_1 J_2}}
\ee
by taking the overlap between the initial and final state in the
string bit theory. The calculation is a (suspiciously) exact
copy of that in the gauge theory: the overlap between the in- and
out-state involves the sum
\be
\qquad \qquad
\sum_{k,l=1}^{J_1} e^{2\pi i (l-k) (p/J-q/J_1)} =
\, {\sin^2 (\pi p \, J_1/ J) \over \pi ^2 (p/J-q/J_1)^2}
\ee
(assuming $J_1$ is large), while
\be
\Delta_p \! - \Delta_{q} = \lambda {(p^2 - q^2{J^2/ J_1^2})}.
\ee
After accounting for normalization factors of $\sqrt{J}$ and such, one obtains
the exact same result as in \cite{seven}, eqn (5.10).

As a perhaps somewhat more instructive example, let us compute
(from the string bit Hamiltonian) the anomalous dimension $\Delta_p$
of $O^J_p$ to second order in the effective string coupling $g_\two$.
To zeroth order we have $H_0 | \, p \, \rangle = (\Delta^{{}^{\! (0)}}\! \! -J)
|\, p\, \rangle$. Next we note that when acting on the space of one-string
and two-string states, the total Hamiltonian takes the form
\be
\left(\! \begin{array}{cc}
H_0 + g_\two^2 \, V_2  \! & g_\two V_{1}\\[1.2mm] g_\two V^*_{1}\! & H_0+ g_\two^2\, V_2
\end{array}
\! \right)\left(\! \begin{array}{c} \! \!\!\! |\, p\, \rangle_{{}_{\!J}}
\\[1.2mm] |\, q\, \rangle_{{}_{\! J_{{}_{\! 1}}\! J_{{}_{\! 2}}}}
\end{array}
\!\!\!\! \right).
\ee
So we find
\be
\label{delta2}
\Delta_p = \Delta_p^{{}^{\! (0)}} + \, g_\two^2\, \Delta_p^{{}^{\! (2)}}\, ,
\qquad \quad
\Delta_p^{{}^{\! (2)}} = \, \langle \, p \, |\, V_2\, | p \, \rangle \; + \,
\sum_q
{|\langle \, p\, |\, V_1\, | q \, \rangle|^2 \over {\Delta^{{}^{\! (0)}}_p \!-\!
\Delta^{{}^{\! (0)}}_q}}
\ee
Inserting (\ref{def}), we can rewrite the last term as
\be
\label{same}
\sum_q
({\Delta^{{}^{(0)}}_p \!\!-\! \Delta^{{}^{(0)}}_q})
|\langle\, p\, |\, \Sigma \, |\, q\, \rangle|^2 .
\ee
This contribution was evaluated in \cite{seven}, with the end-result
(5.17). The first term in (\ref{delta2}), upon inserting
(\ref{drie}), can
be written as \footnote{Here ${\rm h.c.}$ is in fact short hand for three
additional terms: a term $\widetilde{Q}_0 \Sigma \, \widetilde{S}_1$
plus the two hermitian conjugate terms. Further, to obtain (\ref{last}), we
used that $Q$ and $\Sigma^2$ commute when placed between two single string
states, since $\langle \, 1 \,| \Sigma^2 | \, 1 \, \rangle = \sum_{mn}
 \langle \, 1 \,| \Sigma_{mn}^2 | \, 1 \, \rangle$.}
\be
\label{last}
\langle \, p  \, |\, V_2\, | p\,  \rangle = \langle p\, |\,( Q_0^\da\, \Sigma\;
S_1^{\, \da} \, + {\rm h.c.})\, |p\, \rangle
+\, \langle\, p\, |\, V_1\, \Sigma \; |\, p\, \rangle
\ee
(no sum over $\da$).

We can compare this expression with the analysis in section 3.3 of
\cite{seven}. Since $Q^\da$ in the first term directly acts on the
single string state, it represents an nearest neighbor hopping term. We
thus recognize the first term as representing the semi-nearest neighbor interactions,
while the second term on the right-hand side is the non-nearest neighbor
interaction (see figs 8 and 9 in \cite{seven}). The semi-nearest neighbor
interactions turn out to cancel due to anti-symmetry. Upon inserting
a complete set of intermediate states into the second term in (\ref{last}),
one quickly verifies that it takes the exact same form as the contribution
(\ref{same}), again in accordance with the results in \cite{four}
\cite{seven}. We thus
conclude that the total shift in the anomalous dimension is {\it twice}
(\ref{same}), and thus twice (5.17) in \cite{seven}.\footnote{
Our answer thus differs by a factor of two from \cite{seven}. Our
interpretation,
however, is that to compute the total shift in dimensions, one should
diagonalize
the full light-cone Hamiltionian, leading to a mixing between one and
two string
states. This may look at odds with an S-matrix philosophy, where one and
two string
states would not mix. However, both from the gauge theory and string theory
perspective,  there's no reason such mixing would not occur, since
strings with
non-zero $p_+$ never truly separate from each other in the asymptotic region.}

\newsubsection{$J^2/N$ versus $g_s$}

A remaining puzzle is to explain why the effective string coupling
needs to be identified with $g_\two = J^2/N$, rather than with
the fundamental string coupling
\be
\label{renorm}
g_s = {g_2 \lambda^2/ J^2}
\ee
predicted by the AdS/CFT dictionary. An intuitive explanation is that
strings in the plane wave background are in effect confined to a finite
transverse volume, which amounts to an effective dimensional reduction to
two dimensions and a corresponding rescaling of the Newton constant
\cite{seven}. It is important, however, to understand this renormalization
also directly from the worldsheet perspective.\footnote{Ideas similar to
the following discussion were put forward in \cite{rajesh}.}

The physical meaning of the dimensionless parameter $\lambda$ is that
it represents the (inverse) size of the string bit relative to the
worldsheet mass-scale $\mm$ created by the plane wave background.
Calling the string bit size $\ell$, we have
\be
\label{ratio}
\mm\, \ell = 1/\lambda.
\ee
The SYM perturbative regime therefore corresponds to the limit where
the string bit size is much larger than the length scale set by $\mm$.
There is no reason, however, certainly not within the string bit model,
to assume that $\lambda$ is small.

In the limit of large $\lambda$ and large $J$, there will be an
intermediate regime of worldsheet scales $L$ at which the string bits
look infinitesimal, while the mass perturbation $\mm$ is still small.
In this regime, the worldsheet dynamics should look close
to that of strings in flat space. In particular, the string interactions
should have an accurate effective description by means of the
matrix string interaction vertex (\ref{intee}). In addition, the
effective string coupling should be equal to $g_s$ rather than $g_\two$.

Let us compare the two effective interaction vertices (\ref{intee})
and (\ref{def}). Both are a product of a pure twist-field (splitting/joining)
interaction and a quadratic expression in the string oscillators.
It indeed looks like it should be possible to continuously connect
the two operators by means of a suitable renormalization group trajectory,
connecting the small $\lambda$ regime (\ref{def}) with the large
$\lambda$ regime (\ref{intee}). This problem is presently under study
in \cite{djh}.

On general grounds, however, one can already deduce
how the effective couplings should be related in the two regimes.
Namely, a pure twist field interaction
$V_{twist} = \sigma \tilde{\sigma} \Sigma \tilde{\Sigma}$
in the massless continuum worldsheet theory has scale dimension 2.
In the massive theory, however, this conformal dimension will depend non-trivially
on the ratio $(\ref{ratio})$: it interpolates from 2 for $\lambda$ large,
to 0 for small $\lambda$, \cite{djh}. Since the typical length scale of
the worldsheet dynamics is of order $J\ell$, one expect that under
the RG flow the twist field gets renormalized with a factor of
$(\mm  J \ell)^2$.
That is
\be
\label{renorm2}
V_{twist}  \ \ \leftrightarrow \; \ \Sigma \, J^2/\lambda^2
\ee
with $\Sigma$ the string bit version (\ref{sigma}) of the twist field.
This explains the renormalization (\ref{renorm}) of the effective
string coupling.

\newsubsection{Conclusion and outlook}

We have presented a simple string bit system that describes
interacting strings in a plane wave background. The model is complete in
the sense that the space-time symmetry algebra closes at finite string
coupling, without the need for higher order terms other than that we have described
here. The BMN conjecture is that the free string bit Hamiltonian correctly
summarizes all planar interactions of ${\cal N}=4$ gauge theory in the BMN limit;
via our model, one is now free to
extend this conjecture to arbitrary orders in the string loop expansion.
While in principle it is possible that extra
terms may (need to) be added to $H$, these are highly constrained by the
requirement that they preserve the space-time supersymmetry algebra,
as well as the existence of a continuum limit. In this sense, such
possible extra terms would seem to amount to a marginal or irrelevant
deformation of the model, rather than a necessary correction term.

What is the regime of validity of the model?
From the interaction Hamiltonian, we find that string interactions
remain weak as long as
\be
{g_\two \lambda\over J} = {{g_{ym} J}\over \sqrt{N}} , \qquad {\rm and} \ \
\ {g_\two \lambda^2\over J^2} = g_s,
\ee
are both sufficiently small. Outside this regime, the continuum limit ceases
to exist and the string worldsheet
description breaks down. The region $g_\two \lambda> J$ has
indeed been identified in \cite{bmn} as the regime where the strings start to
expand into giant gravitons. It seems likely that a matrix (string) theory
type description takes over in this regime. Indeed it would be interesting
to derive or construct a dual $U(J)$ matrix model that reduces to our bit
model in its strong coupling limit.

It should be straightforward to extend the string bit model to the quiver
set-up of \cite{smoose}, which should in particular clarify the relation with
matrix string theory.
The central role of the permutation group further
suggests that similar ideas as presented here may be used to construct
an interacting string bit theory in the Penrose limit of $AdS^3$.

\newsubsection{Acknowledgements}

It is a pleasure to thank Nissan Itzhaki, Lubos Motl, Sunil Mukhi,
John Pearson, Mukund Rangamani, Marc Spradlin, Diana Vaman and
Erik Verlinde for helpful discussions. This research was supported
by NSF grant 98-02484.

\medskip

\newsubsection{Note added:}

Since the time of writing this paper, several further developments
led to a more precise understanding of both the gauge theory and
string theory side of the story. These new insights made clear
that the original proposal as formulated here needs some
modification.  To clarify the present status of our results, a
brief summary of this subsequent history will be helpful.

Initially, the interaction vertex as proposed in eqn (25)
coincided not only with the reported values of the gauge theory
amplitudes \cite{four}\cite{seven} but also with the three-point
function in string field field theory, as reported in the original
version of \cite{svtwo}. Soon after, however, more careful study
of the gauge theory amplitudes \cite{bkpss}\cite{boston2}, (in
part prompted by our suggestion that operator mixing must in fact
be taken into account) resulted in a new and different answer for
the three-point function, as well as for the shift in the
conformal dimension of the two-impurity states
\cite{bkpss}\cite{boston2}. Simultaneously, it was pointed out
\cite{Pankiewicz:2002gs} that the original string theory
calculation of \cite{svtwo} contained a subtle sign error, and
needed correction as well.\footnote{An alternative construction of
the string interaction vertex, which in fact bears close
resemblance to the vertex (25), was proposed in
\cite{Chu:2002eu}\cite{DiVecchia:2003yp}.}

\newcommand{\gtt}{{{}^{{}_ {>}}}}
\newcommand{\stt}{{{}^{{}_{<}}}}

In \cite{vv}, the following refinement of the bit string vertex
(25) was proposed \be \label{news} S_1^\da =  [
\widehat{Q}^\da_0,\Sigma]\, , \qquad \qquad \widehat{Q}_0 =
Q^\stt_0\!\!-\!Q_0^\gtt \, , \ee where $Q_0 = Q^\stt_0  +
Q^\gtt_0$ is the free supercharge of the bit string theory and the
superscripts indicate the projection onto the term with fermionic
creation ($<$) or annihilation ($>$) operators only. This vertex
was shown to reproduce all known gauge theory amplitudes
\cite{bkpss}\cite{boston2}. In addition, a basis transformation
was proposed that relates the gauge theory basis of single and
multi-trace operators with the string theory basis of single and
multi-string states. This proposed dictionary was verified to
linear order in $g_2$ in \cite{Pearson:2002zs}, where a first
precise match between the gauge theory amplitudes and string
theory interactions was found. This dictionary was independently
proposed in \cite{gmr} and \cite{Gomis:2002wi}, and verified to
second order in \cite{Roiban:2002xr}. Finally, in
\cite{Beisert:2002ff} an effective quantum mechanical model for
capturing the gauge theory interactions was constructed, which was
shown in \cite{Spradlin:2003bw} to be in accord with the string
bit vertex (\ref{news}). Also our general approach of identifying
string scattering amplitudes with two-point functions of
multi-trace operators, rather than multi-point functions, has been
supported by several later results.

A remaining challenge is to construct the continuum limit of the
bit theory. Some aspects of this problem, including (ways of
avoiding) fermion doubling, have been analyzed in
\cite{zhou}\cite{Bellucci:2003qi}\cite{Danielsson:2003yc}.


\end{document}